\documentstyle[epsf,12pt]{article}

\begin{document}

%%%%%%%%%%%%%%%%%%%%%%%%%%%%%%%%%%%%%%%%%%%%%%%%%%%%%%%%%%%%%%
\catcode`@=11
% Redefine caption to put text and formulas in smaller font
\long\def\@caption#1[#2]#3{\par\addcontentsline{\csname
  ext@#1\endcsname}{#1}{\protect\numberline{\csname
  the#1\endcsname}{\ignorespaces #2}}\begingroup
    \small
    \@parboxrestore
    \@makecaption{\csname fnum@#1\endcsname}{\ignorespaces #3}\par
  \endgroup}
\catcode`@=12
%%%%%%%%%%%%%%%%%%%%%%%%%%%%%%%%%%%%%%%%%%%%%%%%%%%%%%%%%%%%
%
%
\newcommand{\newc}{\newcommand}
\newc{\gsim}{\lower.7ex\hbox{$\;\stackrel{\textstyle>}{\sim}\;$}}
\newc{\lsim}{\lower.7ex\hbox{$\;\stackrel{\textstyle<}{\sim}\;$}}
\newc{\gev}{\,{\rm GeV}}
\newc{\mev}{\,{\rm MeV}}
\newc{\ev}{\,{\rm eV}}
\newc{\kev}{\,{\rm keV}}
\newc{\tev}{\,{\rm TeV}}
\newc{\mz}{m_Z}
\newc{\mpl}{M_{Pl}}
\newc{\chifc}{\chi_{{}_{\!F\!C}}}
\newc\order{{\cal O}}
\newc\CO{\order}
\newc\CL{{\cal L}}
\newc\CY{{\cal Y}}
\newc\CH{{\cal H}}
\newc\CM{{\cal M}}
\newc\CF{{\cal F}}
\newc\CD{{\cal D}}
\newc\CN{{\cal N}}
\newc{\eps}{\epsilon}
\newc{\re}{\mbox{Re}\,}
\newc{\im}{\mbox{Im}\,}
\newc{\invpb}{\,\mbox{pb}^{-1}}
\newc{\invfb}{\,\mbox{fb}^{-1}}
\newc{\yddiag}{{\bf D}}
\newc{\yddiagd}{{\bf D^\dagger}}
\newc{\yudiag}{{\bf U}}
\newc{\yudiagd}{{\bf U^\dagger}}
\newc{\yd}{{\bf Y_D}}
\newc{\ydd}{{\bf Y_D^\dagger}}
\newc{\yu}{{\bf Y_U}}
\newc{\yud}{{\bf Y_U^\dagger}}
\newc{\ckm}{{\bf V}}
\newc{\ckmd}{{\bf V^\dagger}}
\newc{\ckmz}{{\bf V^0}}
\newc{\ckmzd}{{\bf V^{0\dagger}}}
\newc{\X}{{\bf X}}
\newc{\bbbar}{B^0-\bar B^0}
\def\bra#1{\left\langle #1 \right|}
\def\ket#1{\left| #1 \right\rangle}
\newc{\sgn}{\mbox{sgn}\,}
\newc{\m}{{\bf m}}
\newc{\msusy}{M_{\rm SUSY}}
\newc{\munif}{M_{\rm unif}}
%
%
%%%%%%%%%%%%%%%%%% Reference Defs %%%%%%%%%%%%%%%%%%
%
%
\def\NPB#1#2#3{Nucl. Phys. {\bf B#1} (19#2) #3}
\def\PLB#1#2#3{Phys. Lett. {\bf B#1} (19#2) #3}
\def\PLBold#1#2#3{Phys. Lett. {\bf#1B} (19#2) #3}
\def\PRD#1#2#3{Phys. Rev. {\bf D#1} (19#2) #3}
\def\PRL#1#2#3{Phys. Rev. Lett. {\bf#1} (19#2) #3}
\def\PRT#1#2#3{Phys. Rep. {\bf#1} (19#2) #3}
\def\ARAA#1#2#3{Ann. Rev. Astron. Astrophys. {\bf#1} (19#2) #3}
\def\ARNP#1#2#3{Ann. Rev. Nucl. Part. Sci. {\bf#1} (19#2) #3}
\def\MPL#1#2#3{Mod. Phys. Lett. {\bf #1} (19#2) #3}
\def\ZPC#1#2#3{Zeit. f\"ur Physik {\bf C#1} (19#2) #3}
\def\APJ#1#2#3{Ap. J. {\bf #1} (19#2) #3}
\def\AP#1#2#3{{Ann. Phys. } {\bf #1} (19#2) #3}
\def\RMP#1#2#3{{Rev. Mod. Phys. } {\bf #1} (19#2) #3}
\def\CMP#1#2#3{{Comm. Math. Phys. } {\bf #1} (19#2) #3}
\relax
%
%
%%%%%%%%%%%%%%%%%%%%%%% latex eqn abrev's %%%%%%%%%%%%%%%%%%%%%%%%%%%%
%
%
\def\beq{\begin{equation}}
\def\eeq{\end{equation}}
\def\bea{\begin{eqnarray}}
\def\eea{\end{eqnarray}}
%
%
%%%%%%%%%%%%%%%%%%%%%%% common abrev's %%%%%%%%%%%%%%%%%
%
%
\newc{\ie}{{\it i.e.}}          \newc{\etal}{{\it et al.}}
\newc{\eg}{{\it e.g.}}          \newc{\etc}{{\it etc.}}
\newc{\cf}{{\it c.f.}}
\def\smuon{{\tilde\mu}}
\def\neut{{\tilde N}}
\def\char{{\tilde C}}
\def\bino{{\tilde B}}
\def\wino{{\tilde W}}
\def\higgsino{{\tilde H}}
\def\sneut{{\tilde\nu}}
\def\stau{{\tilde\tau}}
%
%
%%%%%%%%%%%%%%%%%%%% slashed symbols %%%%%%%%%%%%%%%%%%%%%
%
%
\def\slash#1{\rlap{$#1$}/} % slashes a character
\def\Dsl{\,\raise.15ex\hbox{/}\mkern-13.5mu D} %this one can be subscripted
\def\delsl{\raise.15ex\hbox{/}\kern-.57em\partial}
\def\Ksl{\hbox{/\kern-.6000em\rm K}}
\def\Asl{\hbox{/\kern-.6500em \rm A}}
\def\Qsl{\hbox{/\kern-.6000em\rm Q}}
\def\gradsl{\hbox{/\kern-.6500em$\nabla$}}
%
%
%%%%%%%%%%%%%%%%%%% various symbol abbreviations, vev's etc %%%%%%%%%%%
%
%
\def\bar#1{\overline{#1}}
\def\vev#1{\left\langle #1 \right\rangle}
%
%
%%%%%%%%%%%%%%%%%%% end of intro %%%%%%%%%%%%%%%%%%%%%%%%%%%%%%%%%%%%%

\begin{titlepage}
~~
%\begin{flushright}
%UND-HEP-01-K03\\
%August 2001\\
%\end{flushright}
\vskip 2cm
\begin{center}
{\large\bf Updated Implications of the Muon Anomalous Magnetic Moment 
for Supersymmetry}
\vskip 1cm
{\normalsize\bf
Mark Byrne, Christopher Kolda and Jason E.~Lennon} \\
\vskip 0.5cm
{\it Department of Physics, University of Notre Dame\\
Notre Dame, IN~~46556, USA\\[0.1truecm]
}

\end{center}
\vskip .5cm

\begin{abstract}
We re-examine the bounds on supersymmetric particle masses in light of the
 new E821 data on the muon anomalous magnetic moment and the revised
 theoretical calculations of its hadronic contributions. The current
 experimental excess is either $1.5\sigma$ or $3.2\sigma$ depending on
 whether $e^+e^-$ or $\tau$-decay data are 
used in the theoretical calculations for
 the leading order hadronic processes. Neither result is compelling evidence
 for new physics. However, if one interprets the excess as coming from
 supersymmetry, one can obtain upper mass bounds on many of
the particles of the
 minimal supersymmetric standard model (MSSM). Within this framework
 we provide a general analysis of the lightest masses as a function of the
 deviation so that future changes in either experimental data or theoretical
 calculations can easily be translated into upper bounds at the desired level
 of statistical significance.  In addition, we give specific bounds on
 sparticle masses in light of the latest experimental and theoretical
 calculations for the MSSM with universal slepton masses, with and
 without universal gaugino masses.

% We find that the unconstrained MSSM predicts 4 sparticles with masses below
%$585 \gev$ (at the  $1\sigma$ level of the excess) excluding the $\tau$ data
%in the SM analysis. Including the tau decay data leads to the significantly
%less stringent bound of $1 \tev$ on the four lightest sparticles. There is
%no lower bound on$\tan\beta$ implied by the data.
\end{abstract}

\end{titlepage}

\setcounter{footnote}{0}
\setcounter{page}{1}
\setcounter{section}{0}
\setcounter{subsection}{0}
\setcounter{subsubsection}{0}

%%%%%%%%%%%%%%%%%%%%%%%%%%%%%%%%%%%%%%%%%%%%%%%%%%%%%%%%%%%%%%%%%%%%%%%

In February 2001 the Brookhaven E821 experiment~\cite{e821a} reported
evidence for a deviation of the muon magnetic moment, $a_\mu$, from the 
Standard Model expectation of about $2.7\sigma$.
Immediately following that announcement 
appeared a number of papers analyzing the reported excess in terms of
various forms of new physics, including supersymmetry
(SUSY)~\cite{czar,kane,martinwells,others}.
Shortly thereafter, errors in the theoretical
calculation of the magnetic moment within the Standard Model were
discovered. In particular, the sign of the light-by-light hadronic
contribution to $a_{\mu}$ was found to be in error~\cite{lbl}, shifting
the theoretical value by roughly $17\times 10^{-10}$ in the direction
of the E821 data. The resulting discrepancy between data and theory
was then only about $1.6\sigma$, leaving little indication of new physics.

Since that initial rise and ebb of interest in $a_\mu$, there has 
been progress both theoretically and experimentally. 
On the theory side, 
besides the recalculation of the light-by-light scattering,
there have also appeared new calculations of the other 
hadronic contributions to $a_\mu$~\cite{narison,ty,davier2,hagiwara2}.
On the experimental front, E821 has announced~\cite{e821b} 
an updated measurement of $a_\mu$ using
a data set four times larger than that analyzed in Ref.~\cite{e821a}.
From their combined data sets, they obtain a world average for $a_\mu$
slightly higher than their previous measurement, but with
significantly smaller errors. These various developments have now
placed the discrepancy between experiment and theory in the range
$1.5\sigma$ to $3.2\sigma$.

The current deviation of E821's measurement from the Standard Model 
provides no compelling evidence in favor of
new physics. However, the attention paid to this process over the last
year warrants a re-examination of the bounds that can be placed on new
physics by the current data, in particular, 
on new SUSY particles. If the small deviation in $a_\mu$ {\em
is}\/ a sign for new physics, then
the SUSY explanation is, for many of us, the
most exciting of the various proposals, since it implies SUSY at a mass
scale not far above the weak scale. In particular, it implies a light
slepton and a light gaugino, though ``light'' can still be as heavy as
many hundred GeV.

This paper plays two roles. First and more importantly, we
present general bounds on the spectrum of SUSY masses for any
deviation of the 
muon magnetic moment from the theoretical Standard Model value.  
Any changes in the Standard Model calculation or to the data itself
will not affect this general analysis; when new data or calculations
appear, revised bounds can simply be read off the plots contained
here.  Second, we will determine specific $1\sigma$ upper 
bounds on SUSY states in light of the most recent
E821 data and the latest theoretical calculations. 

The existence of these bounds will rely on very simple and clearly stated
 assumptions about the SUSY particle spectrum; these assumptions will not
 include a fine-tunimg constraints.

\section{SUSY and $a_\mu$}

The measurement performed by the E821 collaboration is of the
muon's anomalous magnetic 
moment~\footnote{See, however Ref.~\cite{edm}.}, which is to say, the
coefficient $a_\mu$ of the non-renormalizable operator 
\beq
\frac{a_\mu}{2m_\mu}\bar\psi\, \sigma^{\alpha\beta}\psi
F_{\alpha\beta}.
\eeq
Within the Standard Model, $a_\mu$ receives
contributions from QED, electroweak and hadronic processes, the latter
usually separated into contributions from  
vacuum polarization and light-by-light scattering. While the
QED and electroweak contributions are well understood, the 
hadronic calculations are under scrutiny and require experimental
input. For the purposes of this paper we will use the most recent 
determination of the hadronic contributions available.

The QED and electroweak contributions to $a_\mu$ are known:
\beq
\begin{array}{llrr}
a_{\mu}^{\rm QED} &=& 11\,658\,470.56\, (0.29) &
\quad\mbox{\cite{qed}}  \\[2truemm] 
a_{\mu}^{\rm EW} &=& 15.2\phantom{2}\, (0.1)\phantom{2} & \mbox{\cite{ew}}
\end{array} 
\eeq
where we are expressing $a_\mu$ in units of $10^{-10}$.
In the hadronic sector, the next-to-leading order (NLO) contributions
have been known for several years, while the and light-by-light
(LBL) contributions have been updated in the last year:
\beq
\begin{array}{llrr}
a_{\mu}^{\rm Had, NLO} &=& -10.0\, (0.6) & \quad\mbox{\cite{nlo}} \\[2truemm]
a_{\mu}^{\rm Had, LBL} &=& 8.0\, (4.0)	& \mbox{\cite{lbl}}.
\end{array}
\eeq
The calculation of leading order (LO) 
hadronic contributions requires input from
experiment. Davier \etal~\cite{davier2} have calculated the LO
hadronic pieces using $e^+e^-$ scattering data and alternatively with
a combination of $e^+e^-$ and $\tau$-decay data. They find:
\beq
a_{\mu}^{\rm Had, LO} = \left\lbrace
\begin{array}{rl}
684.7\, (7.0) & \mbox{(no $\tau$ data)}\\
701.9 \,(6.2) & \mbox{($\tau$ data).} 
\end{array} \right.
\eeq
Hagiwara~\etal\ have also calculated the LO hadronic contribution  
without the $\tau$ data in Ref.~\cite{hagiwara2}:
\beq
\begin{array}{llrl}
a_{\mu}^{\rm Had, LO} &=& 683.1\, (6.2) & \mbox{(no $\tau$ data)}\\
\end{array} 
\eeq
which is consistent with the above evaluation. For our estimate of $a_{\mu}$ without $\tau$
 data we will
use the average of the two results stated above and the larger error.
Adding the contributions gives the standard model prediction for $a_{\mu}$: 
\beq
a_{\mu} = \left\lbrace
\begin{array}{rl}
11\,659\,167.7\,(8.1) &  \mbox{(no $\tau$ data)}\\
11\,659\,185.7\,(7.4) &  \mbox{($\tau$ data)}.
\end{array} \right.
\eeq
Since the two values are mutually inconsistent, we will not combine
them into a single prediction of $a_\mu$.
 
The new measurement made by E821 is~\cite{e821b}
$a_\mu^{\rm E821} = 11\,659\,204\,(7)(5)\times 10^{-10}$
yielding a world average of:
\beq a_\mu^{\rm exp} = 11\,659\,203\,(8)\times 10^{-10} \eeq
from which one deduces a discrepancy between the experiment and the
Standard Model of
\beq
\delta a_\mu = \left\lbrace
\begin{array}{lll}
35(11)\times 10^{-10}  &  \mbox{(no $\tau$ data)} \\
17(11)\times 10^{-10}  &  \mbox{($\tau$ data)}  
\end{array} \right.
\label{dev}
\eeq
where we have added the theoretical and experimental errors in
quadrature. Thus, the deviation is either $3.2\sigma$ or $1.5\sigma$, 
depending on whether or not the $\tau$ data is used.
In the analysis that follows, we present results using both values of
$\delta a_\mu$ from Eq.~(\ref{dev}). 
%When the hadronic contributions become
%more well understood, the discrepancy in $a_\mu$ may presumably change.
%When this happens, the numerical bounds on SUSY masses can be read directly
%from the plots contained herein.
%
%When (if) the theoretical picture settles down, the uncertainty in
%the deviation will presumably asymptote to a fixed value (perhaps 
%nonzero) and the numerical implications for the SUSY mass bounds 
%can be read directly from the plots contained herein.   

The SUSY contributions to $a_\mu$ have been known since the early days
of SUSY and have become more complete with time\cite{calcs}. In this
paper we will follow the notation of Ref.~\cite{martinwells} which has the
advantage of using the standard conventions of Haber and
Kane~\cite{haberkane}; any convention which we do not define here can be
found in either of these two papers. 

Prior to the revision of the theoretical calculations and the release
of the newest E821 data, there were a number of analyses in the
context of SUSY~\cite{czar,kane,martinwells,others}.
The present authors also presented an analysis of the full 
minimal supersymmetric standard model (MSSM) with
and without gaugino unification and neutralino LSP
constraints~\cite{ourold}.

%However $1\sigma$ bounds on SUSY partners can still
%be found, though clearly no $2\sigma$ bounds will be possible.
%An analysis of minimal supergravity-motivated models
%using the new experimental and theoretical values
%has also recently appeared~\cite{newnath}. 

\subsection{The diagrams}

In the mass eigenbasis, there are only two one-loop SUSY 
diagrams which contribute to $a_\mu$, shown in Figure~1. The first has
an internal loop of smuons and neutralinos, the second a loop of
sneutrinos and charginos. But the charginos, neutralinos and even the
smuons are themselves admixtures of various interaction eigenstates
and we can better understand the physics involved by working in terms
of these interaction diagrams, of which there are many more than two.
We can easily separate the leading and sub-leading diagrams in the
interaction eigenbasis by a few simple observations.

%%%%%%%%%%
\begin{figure}
\centering
\epsfxsize=4.25in
\hspace*{0in}
\epsffile{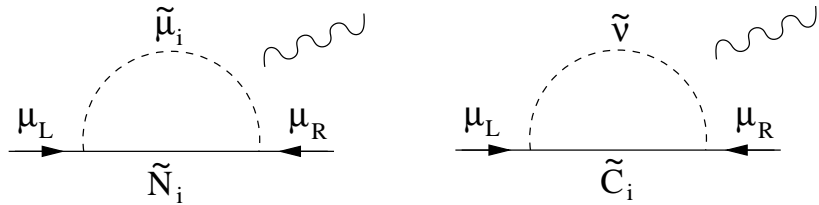}
\caption{Supersymmetric diagrams contributing to $a_\mu$ at one-loop.}
\label{simple}
\end{figure}
%%%%%%%%%%%

First, the magnetic moment operator is a helicity-flipping
interaction. Thus
any diagram which contributes to $a_\mu$ must involve a helicity flip
somewhere along the fermion current. This automatically divides the
diagrams into two classes: those with helicity flips on the external
legs and those with flips on an internal line. For those in the first
class, the amplitude must scale as $m_\mu$; for those in the second,
the amplitudes can scale instead by $m_{\rm SUSY}$, where $m_{\rm
SUSY}$ represents the mass of the internal SUSY fermion (a chargino or
neutralino). Since $m_{\rm SUSY}\gg m_\mu$, it is the latter class
that will typically dominate the SUSY contribution to
$a_\mu$. Therefore we will restrict further discussion to this latter
class of diagrams alone.

Secondly, the interaction of the neutralinos and charginos with the
(s)muons and sneutrinos occurs either through their higgsino or 
gaugino components. Thus each vertex implies a factor of either
$y_\mu$ (the muon Yukawa coupling) or $g$ (the weak and/or hypercharge
gauge coupling). Given two vertices, the diagrams therefore scale as
$y_\mu^2$, $gy_\mu$ or $g^2$.
In the Standard Model, $y_\mu$ is smaller than $g$ by
roughly $10^{-3}$. In the minimal SUSY standard model (MSSM) 
at low $\tan\beta$, this ratio is
essentially unchanged, but because $y_\mu$ scales as $1/\cos\beta$, at
large $\tan\beta$ ($\sim60$) the ratio can be reduced to roughly
$10^{-1}$. Thus we can safely drop the $y_\mu^2$ contributions from
our discussions, but at large $\tan\beta$ we must preserve the
$gy_\mu$ pieces as well as the $g^2$ pieces.\footnote{Pieces 
which are dropped from our discussion
are still retained in the full numerical calculation.}

The pieces that we will keep are therefore shown in Figures~2. In
Fig.~2(a)-(e) are shown the five neutralino contributions which scale 
as $g^2$
or $gy_\mu$; in Fig.~2(f) is the only chargino contribution, scaling as
$gy_\mu$. The contributions to $a_\mu$ from the $i$th
neutralino and the $m$th smuon due to each of these component
diagrams are found to be:
\beq
\delta a_\mu = \frac{1}{48\pi^2}\frac{m_\mu m_{\neut_i}}{m_{\smuon_m}^2}
F_2^N(x_{im})
\times \left\lbrace \begin{array}{lc}
g_1^2 N_{i1}^2 X_{m1}X_{m2} & (\bino\bino)\\
g_1g_2 N_{i1}N_{i2} X_{m1}X_{m2} & (\wino\bino) \\
-\sqrt{2}g_1y_\mu N_{i1}N_{i3} X_{m2}^2 & (\higgsino\bino) \\
\frac{1}{\sqrt{2}}g_1y_\mu N_{i1}N_{i3} X_{m1}^2 & (\bino\higgsino)\\
\frac{1}{\sqrt{2}}g_2y_\mu N_{i2}N_{i3} X_{m1}^2 & (\wino\higgsino)
\end{array} \right.
\eeq
and for the $k$th chargino and the sneutrino:
\beq 
\delta a_\mu = -\frac{1}{24\pi^2}\frac{m_\mu m_{\char_k}}{m_{\sneut}^2}
F_2^C(x_{k})\,g_2y_\mu U_{k2}V_{k1}.\quad\quad (\wino\higgsino) \eeq
The matrices $N$, $U$ and $V$ are defined in the appendix along with
the functions $F_2^{N,C}$. A careful comparison to the equations in
the appendix will reveal that we have dropped a number of complex
conjugations in the above expressions; it has been shown
previously~\cite{kane,martinwells} that the SUSY contributions to
$a_\mu$ are maximized for real entries in the mass matrices and so we
will not retain phases in our discussion.

%%%%%%%%%%
\begin{figure}
\centering
\epsfxsize=4.25in
\hspace*{0in}
\epsffile{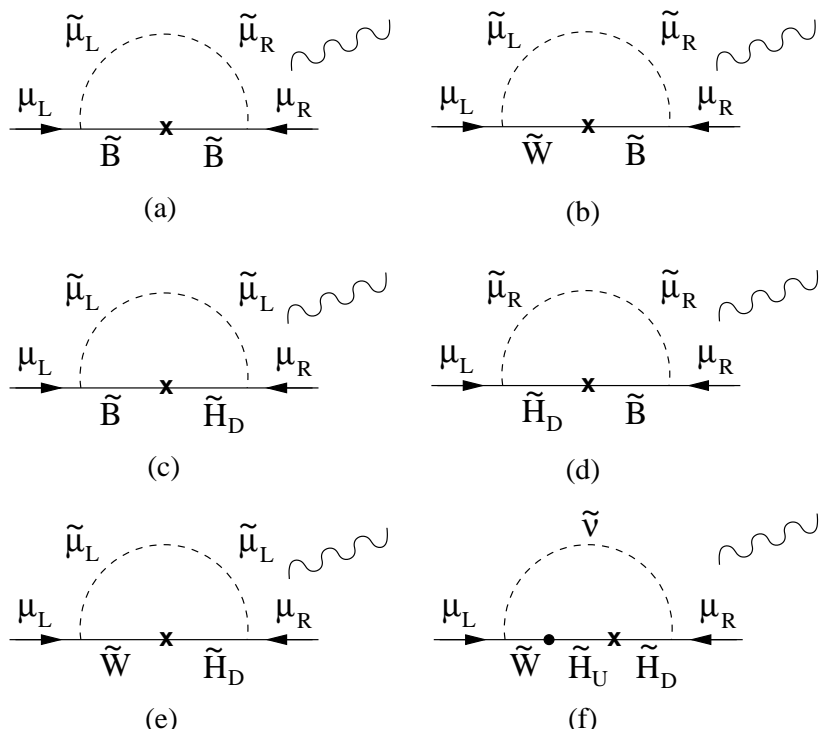}
\caption{Diagrams contributing to $a_\mu$ in the interaction eigenbasis.
}
\label{intdiags}
\end{figure}
%%%%%%%%%%%

In many of the previous analyses of the MSSM parameter space, it was
found that it is the chargino-sneutrino
diagram at large $\tan\beta$ 
that can most easily generate values of $\delta a_\mu$ large
enough to explain the observed discrepancy.
From this observation, one can obtain an upper mass bound on the
lightest chargino and the muon sneutrino. However, this behavior
is not completely generic. For example,
Martin and Wells emphasized that the $\bino\bino$
neutralino contribution could {\it by itself}\/ be large enough to
generate the old observed excess in $a_\mu$, and since it has no
intrinsic $\tan\beta$ dependence, they could explain the old excess with
$\tan\beta$ as low as 3. We can reproduce their result in a simple way
because the $\bino\bino$ contribution has a calculable upper bound at which
the smuons mix at $45^\circ$, $m_{\smuon_1}\ll m_{\smuon_2}$, and
$m_{\neut_1}\ll m_{\neut_{2,3,4}}$ with $\neut_1=\bino$. Then
\beq\left|\delta a_\mu\right|_{(\bino\bino)}
\leq \frac{g_1^2}{32\pi^2}\frac{m_\mu
m_{\neut_1}}{m_{\smuon_1}^2}\simeq
3800\times10^{-10}\times \left(\frac{m_{\neut_1}}{100\gev}\right)
\left(\frac{100\gev}{m_{\smuon_1}}\right)^2\eeq
where we have used the fact that $(X_{11}X_{12})\leq\frac12$ and
$F_2^N\leq 3$ and have included a 7\% two-loop suppression factor.
Though any real model will clearly suppress this
contribution somewhat, this is still $10^2$ times larger than needed
experimentally. 

This pure $\bino\bino$ scenario is actually an experimental 
worst-case, particularly for hadron colliders. The only
sparticles that are required to be light are a single neutralino
(which is probably $\bino$-like) and a single $\smuon$. The neutralino
is difficult to produce, and if stable, impossible to detect
directly. The neutralino could be indirectly observed in the decay of
the $\smuon$ as missing energy, but production of a $\smuon$ at a
hadron machine is highly suppressed. In the worst of all possible
worlds, E821 could be explained by only these two light sparticles,
with the rest of the SUSY spectrum hiding above a TeV. Further, even
the ``light'' sparticles can be too heavy to produce at a $500\gev$
linear collider. While this case
is in no way generic, it demonstrates that the E821 excess 
does not provide any sort of no-lose theorem (even at $1\sigma$) 
for the Tevatron, the LHC or the NLC.

This raises an important experimental question: how many of the MSSM 
states must be ``light'' in order to explain the E821 data? In the
worst-case, it would appear to be only two.
Even in the more optimistic scenario in which the chargino diagram
dominates $\delta a_\mu$, the answer naively appears to be two: 
a single chargino and a single sneutrino. In this limit, 
\beq\left|\delta a_\mu\right|_{(\char\sneut)}
\leq \frac{g_2 y_\mu}{24\pi^2}\frac{m_\mu m_{\char_1}}{m_\sneut^2}
\left|F^C_2\right|_{\rm max} \lsim 2600\times 10^{-10}\times
\left(\frac{m_{\char_1}}{100\gev}\right)
\left(\frac{100\gev}{m_{\sneut}}\right)^2
\left(\frac{\tan\beta}{30}\right)\eeq
where we have bounded $|F_2^C|$ by 10 by assuming $m_\sneut\lsim 1\tev$.
But this discussion is overly simplistic, as we will see.

\subsection{Mass correlations}

There are a total of 9 separate sparticles which can enter the
loops in Fig.~\ref{simple}: 1 sneutrino, 2 smuons, 2 charginos and 4
neutralinos. The mass spectrum of these 9 sparticles is determined
entirely by 7 parameters in the MSSM: 2 soft slepton masses($m_L$,
$m_R$), 2 gaugino masses ($M_1$, $M_2$), the $\mu$-term, a soft
trilinear slepton coupling ($A_\smuon$) and finally $\tan\beta$. Of
these, $A_\smuon$ plays almost no role at all and so we leave it out of
our discussions (see the appendix). And in some well-motivated
SUSY-breaking scenarios, $M_1$ and $M_2$ are also correlated. Thus
there are either 5 or 6 parameters responsible for setting 9 sparticle
masses. There are clearly non-trivial correlations among the
masses which can be exploited in setting mass limits on the sparticles.

First, there are well-known correlations between the
chargino and neutralino masses; for example, 
a light charged $\char_i\sim\wino$
implies a light neutral $\neut_j\sim\wino$ and vice-versa. 

There are also
correlations in mixed systems (\ie, the neutralinos, charginos and
smuons) between the masses of the eigenstates and the size of their
mixings. Consider the case of the smuons in particular; their mass
matrix is given in the appendix. On diagonalizing, the left-right
smuon mixing angle is given simply by:
\beq\tan 2\theta_\smuon\simeq\frac{2 m_\mu\mu\tan\beta}{M_L^2-M_R^2}.\eeq
The chargino contribution is maximized for large smuon mixing and
large mixing occurs when the numerator is of order or greater 
than the denominator; since the former is suppressed by $m_\mu$, one
must compensate by having either a very large $\mu$-term in the numerator or 
nearly equal $M_L$ and $M_R$ in the denominator, both of which have
profound impacts on the spectrum.

%There is one more type of correlation that is very useful at large
%$\tan\beta$. It has been said that the chargino contribution to
%$\delta a_\mu$ implies a single light chargino and a single light
%sneutrino (plus the light smuon demanded by SU(2) to be the partner of
%the sneutrino).But in fact, $a_\mu$ also puts upper bounds on the
%heavier chargino. The basic reasoning is simple. The chargino diagram
%requires significant $\wino\higgsino$ mixing. But
%in the chargino mass matrix, the off-diagonal elements are fixed at the
%$W$-mass scale, so large mixing requires $M_2^2-\mu^2\sim m_W^2$. By
%further requiring that one of the eigenvalues be light one realizes 
%immediately that the heavier eigenvalue must also be relative
%light. (Certainly it is easy to see that in the limit in which one
%chargino is light and the other is infinitely massive, the mixing
%would go to zero.) Thus we should be able to obtain some kind of upper
%bound on the mass of $\char_2$ from the E821 data, at least at large
%$\tan\beta$. Further, since light charginos imply small $M_2$ and
%$\mu$, then bounds probably also exist on the masses
%of at least 3 of the neutralinos. 

There is one more correlation/constraint that we feel is natural to
impose on the MSSM spectrum: slepton mass universality. It is
well-known that the most general version of the MSSM produces huge
flavor-changing neutral currents (FCNCs) unless some external order is
placed on the MSSM spectrum. By far, the simplest such order is for
sparticles with the same gauge quantum numbers to be
degenerate. This requirement is most stringent in the squark sector,
but also holds in the slepton sector due to non-observation of $\mu\to
e\gamma$ and $\tau\to\mu\gamma$. In any case, mechanisms which
generate degeneracy in the squarks usually do so among the sleptons as well.
Thus
we assume $m_{\stau_L}=m_{\smuon_L}\equiv m_L$ and
$m_{\stau_R}=m_{\smuon_R}\equiv m_R$. 
Then the mass matrix for the stau sector
is identical to that of the smuons with the replacement $m_\mu\to
m_\tau$ in the off-diagonal elements. This enhancement of the mixing
in the stau sector by $m_\tau/m_\mu\simeq 17$ implies that
$m_{\stau_1} < m_{\smuon_1}$. In particular, if 
\beq M_L^2 M_R^2<m^2_\tau\,\mu^2\tan^2\beta\eeq
then $m^2_{\stau_1}<0$ and QED will be broken by a stau vev. Given
slepton universality, this imposes a constraint on the smuon mass
matrix:
\beq M_L^2 M_R^2 > \left(\frac{m_\tau}{m_\mu}\right)^2 m_\mu^2\,\mu^2
\tan^2\beta\eeq
or on the smuon mixing angle:
\beq\tan2\theta_\smuon < \left(\frac{m_\mu}{m_\tau}\right)
\frac{2M_LM_R}{M_L^2-M_R^2}\eeq
where $M_{L,R}$ are the positive roots of $M^2_{L,R}$. While not
eliminating the possibility of $\theta_\smuon\simeq 45^\circ$, 
this formula shows
that a fine-tuning of at least 1 part in 17 is needed to obtain
$\CO(1)$ mixing. We will not apply any kind of fine-tuning
criterion to our analysis, yet we will find that this slepton mass
universality constraint sharply reduces the upper bounds on slepton
masses which we are able to find in our study of points in MSSM
parameter space.

(As an aside, if one assumes slepton mass universality at some
SUSY-breaking messenger scale above the weak scale, 
Yukawa-induced corrections will
break universality by driving the stau masses down. This effect would
further tighten our bounds on smuon masses and mixings.)

The above discussion has an especially large impact on the worst-case
scenario in which the $\bino\bino$ contributions dominates $\delta a_\mu$.
For generic points in MSSM parameter space,
one expects that $\tan 2\theta_\smuon\lsim1/17$ which reduces the
size of the $\bino\bino$ contribution by a factor of 17. As a
byproduct, the masses required for explaining the E821 anomaly are
pushed back toward the experimentally accessible region.

\section{Numerical results}

Now that we have established the basic principle of our analysis, we
will carry it out in detail. We will concentrate on three basic
cases. The first case is the one most often considered in the literature:
gaugino mass unification. Here one assumes that the weak-scale gaugino
mass parameters ($M_1$ and $M_2$) are equal at the same scale at which
the gauge couplings unify. This implies that at the weak scale
$M_1=(5/3)(\alpha_1/\alpha_2)M_2$. The second case we consider is
identical to the first with the added requirement that the lightest
SUSY sparticle (LSP) be a neutralino. This requirement is motivated by
the desire to explain astrophysical dark matter by a stable
LSP. Finally we will also consider the most general case in which all
relevant SUSY parameters are left free independent of each other; we
will refer to this as the ``general MSSM'' case.

The basic methodology is simple: we put down a
logarithmic grid on the space of MSSM parameters ($M_1$, $M_2$, $m_L$,
$m_R$ and $\mu$) for several choices of $\tan\beta$. The grid extends
from $10\gev$ for $M_1$, $M_2$ and $\mu$, and from $50\gev$ for
$m_{L,R}$, up to $2\tev$ for all mass parameters. For the case in which
gaugino unification is imposed $M_2$ is no longer a free parameter and
our grid contains $10^8$ points. For the general MSSM case
our grid contains $3\times10^9$ points. Only $\mu>0$ is considered since
that maximizes the value of $\delta a_\mu$. Finally, for our limits on
$\tan\beta$ we used an adaptive mesh routine which did a better job of
maximizing $\delta a_\mu$ over the space of MSSM inputs. By running
with grids  of varying resolutions and offsets we estimate the error
on our mass bounds to be less than $\pm 5\%$.

\subsection{Bounds on the lightest sparticles}

Perhaps the most important information that can be garnered from the
E821 data is an upper bound on the scale of sparticle masses. In
particular, one can place upper bounds on the masses of the lightest
sparticle(s) as a function of $\delta a_\mu$. 
%Previous analyses have
%often followed this approach, deriving upper bounds on the lightest
%sparticle from among the gauginos and sleptons, or even more
%specifically, from among the charginos and smuons. We too derive
Here we will derive bounds on the lightest slepton and chargino, but
we will also derive bounds on the lightest several sparticles, independent of
their identity.

These bounds on additional light sparticles provide an important
lesson. Without them there remains the very real possibility that the
E821 data is explained by a pair of light sparticles and that the
remaining SUSY spectrum is out of reach experimentally. But our
additional bounds
will give us some indication not only of where we can find SUSY, but
also of how much information we might be able to extract about the
fundamental parameters of SUSY --- the more sparticles we detect and
measure, the more information we will have for disentangling the
soft-breaking sector of the MSSM.

\begin{figure}
\centering
\parbox{3.05truein}{
\epsfxsize=3in
\hspace*{0in}
\epsffile{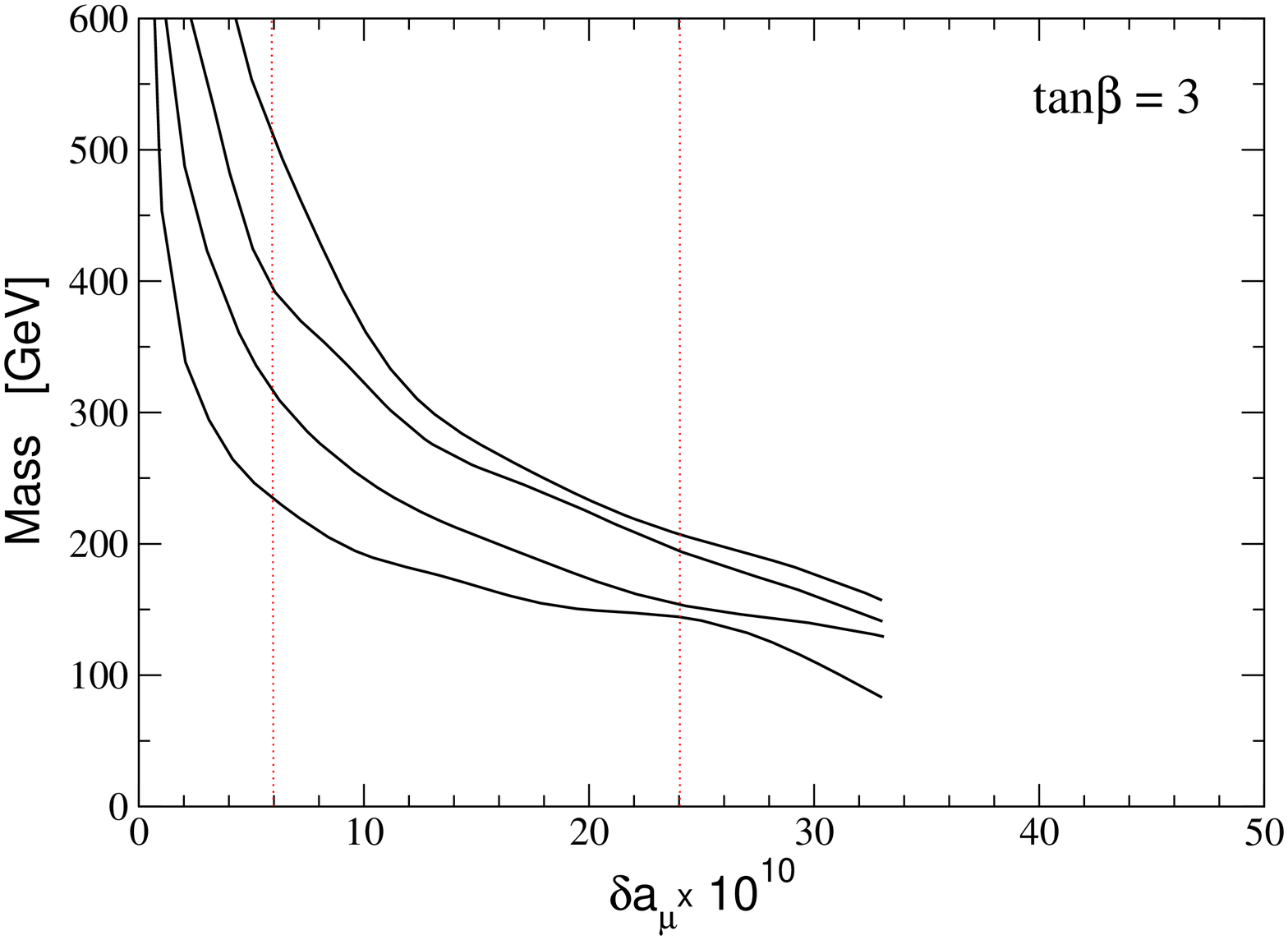}
}~\parbox{3.05truein}{
\epsfxsize=3in
\hspace*{0in}
\epsffile{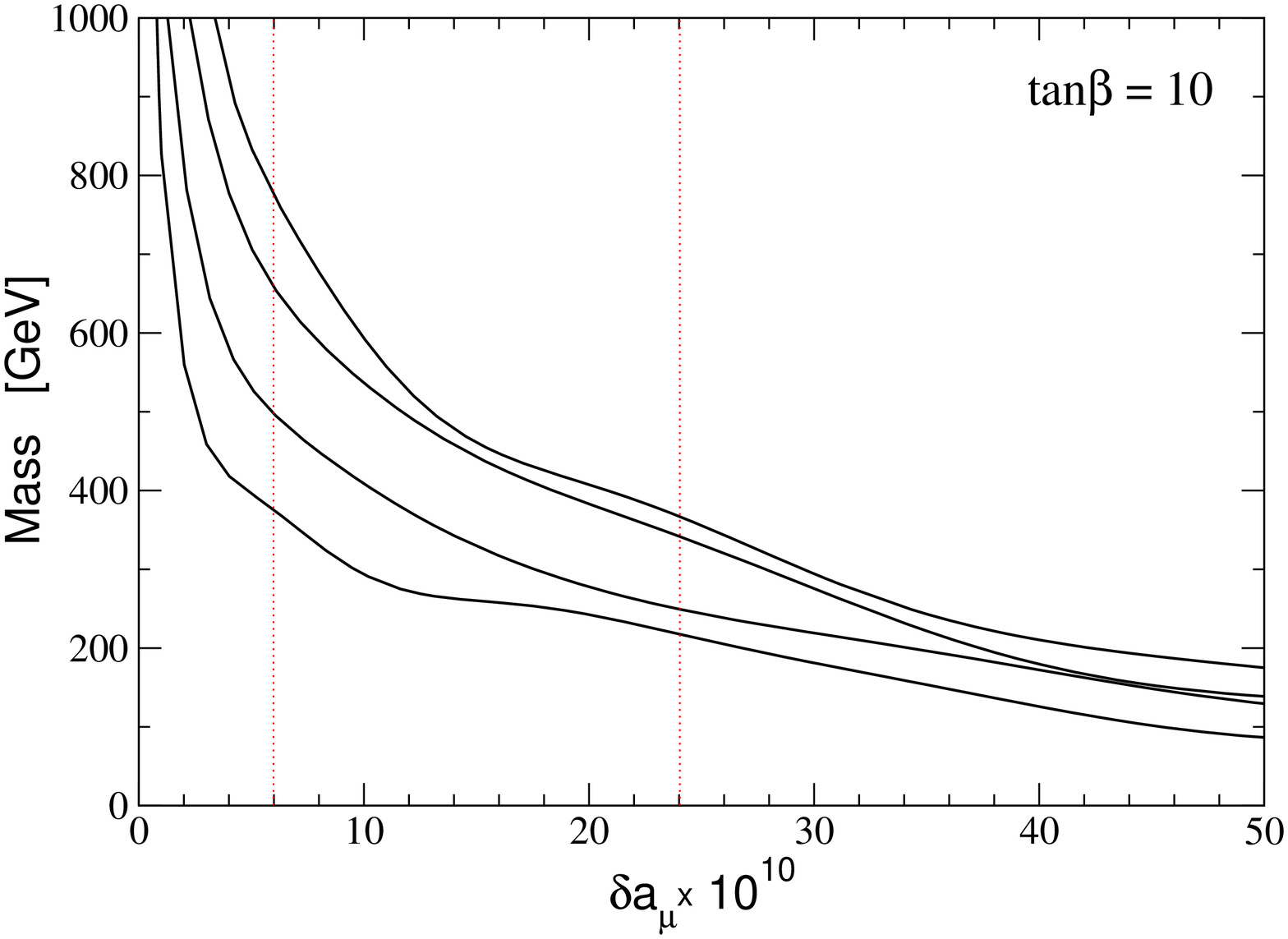}
}
\parbox{5truein}{~~~\\~~~~\\}
%\vspace{0.5truein}
\parbox{3.05truein}{
\epsfxsize=3in
\hspace*{0in}
\epsffile{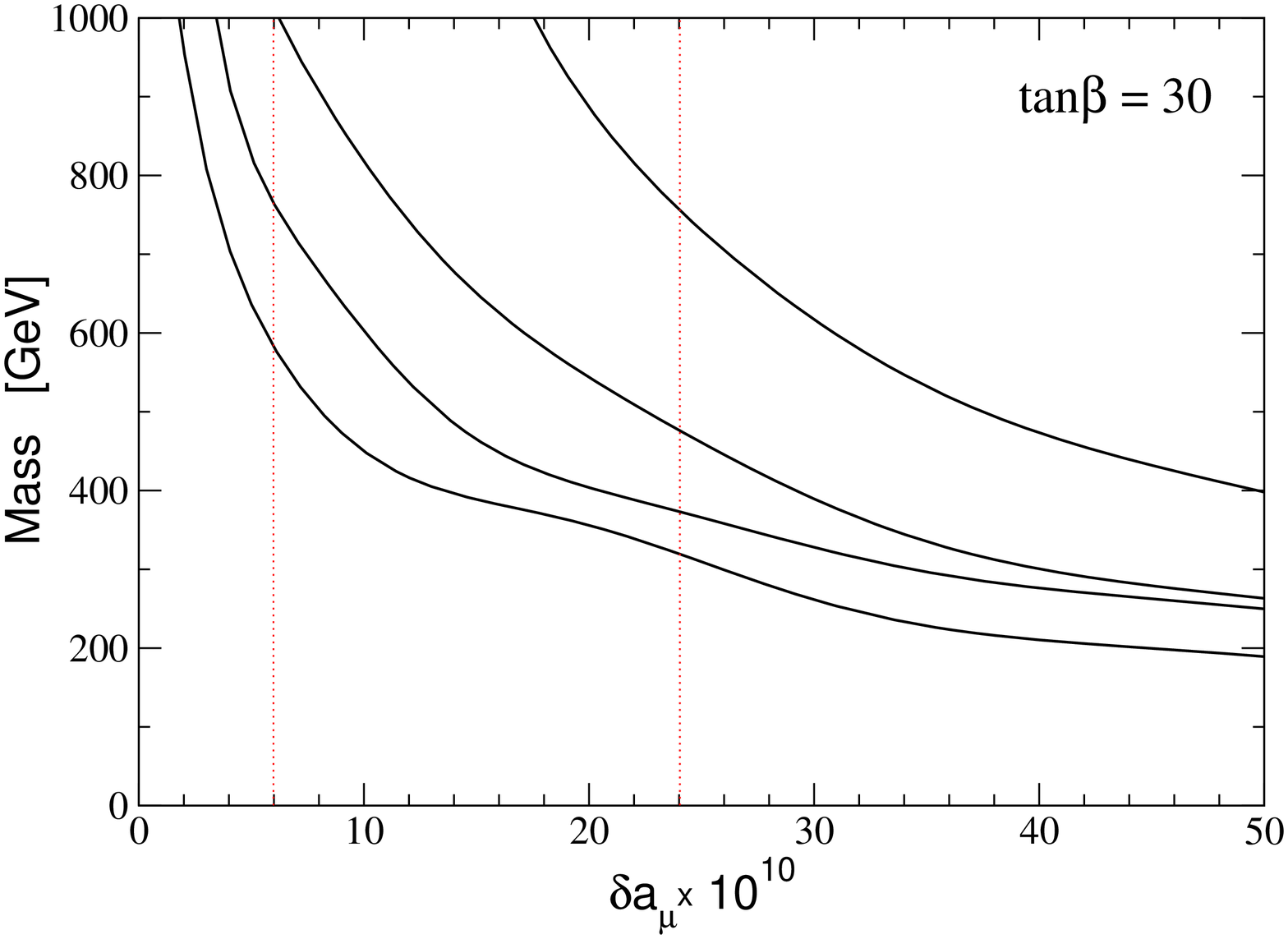}
}~\parbox{3.05truein}{
\epsfxsize=3in
\hspace*{0in}
\epsffile{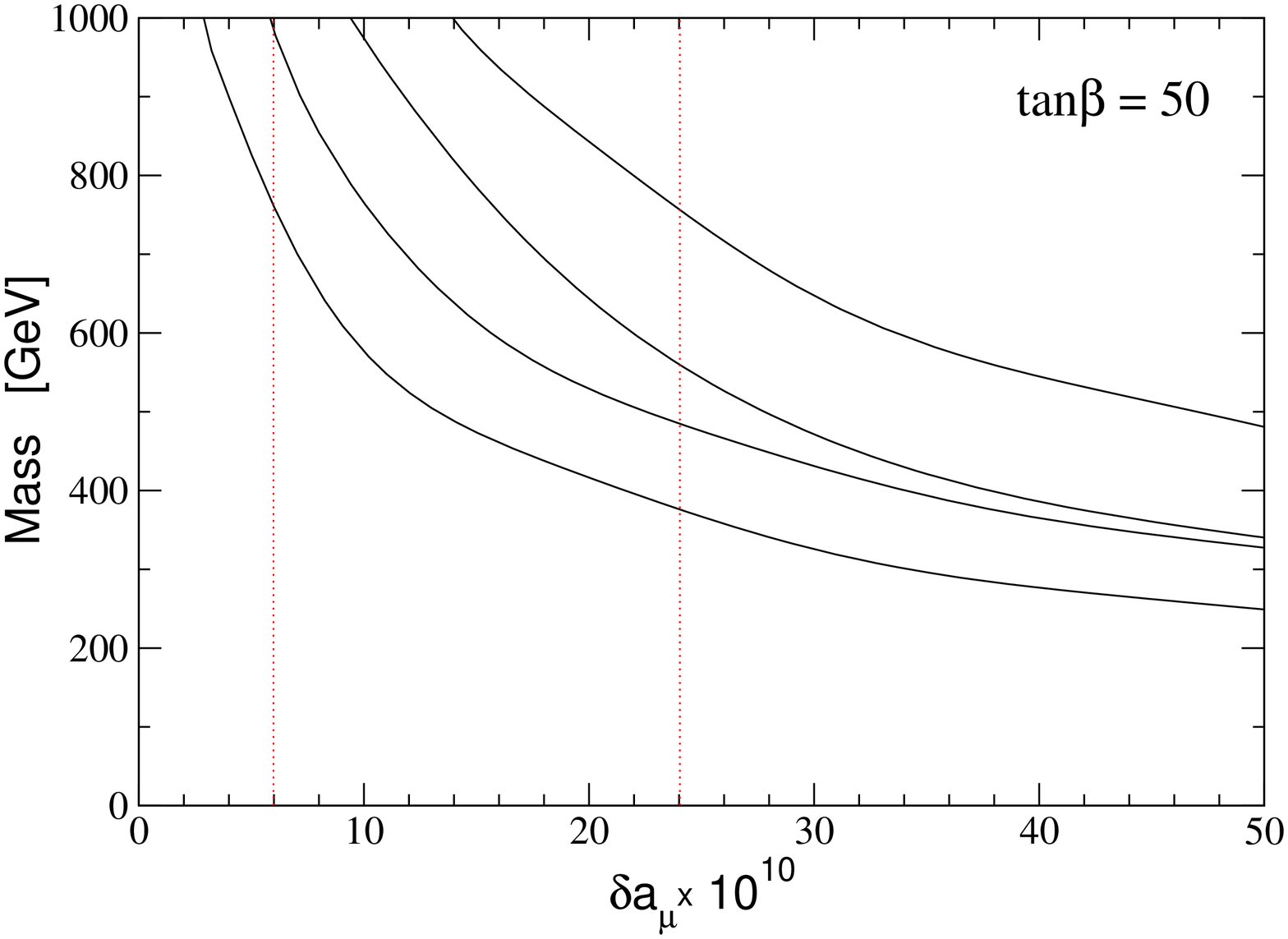}
}
\caption{Bounds on the masses of the four lightest sparticles as a
function of $\delta a_\mu$ for $\tan\beta=3$, 10, 30 and 50.
These figures assume gaugino unification only. The vertical dotted
lines represent the $1\sigma$ 
bounds using the $\tau$-decay data
($\delta a_{\mu} = 6$) or not
($\delta a_{\mu} = 24$).}
\label{massfig}
\end{figure}

In Fig.~\ref{massfig} we have shown the upper mass bounds for the
lightest four sparticles assuming gaugino mass unification. 
These bounds are not bounds on individual species of sparticles (which
will come in the next section and always be larger than these bounds)
but simply bounds on whatever sparticle happens to be lightest.
The important points to note are: {\it (i)}\/ the maximum values of
the mass correspond to the largest value of $\tan\beta$, which is to
be expected given dominance of the chargino diagram at large
$\tan\beta$; {\it (ii)}\/ the $1\sigma$ limit without (with) the $\tau$ 
data requires
at least 2 sparticles to lie below roughly $490\,(990)\gev$;
{\it (iii)} for the central value of the E821 data, at least 4
sparticles must lie below $585\,(915)\gev$;
and {\it (iv)}\/ for low values of $\tan\beta$ a maximum
value of $\delta a_\mu$ is reached (we will return to this later).

%%%%%%%%%%
%%%%%%%%%%%

\begin{figure}
\centering
\epsfxsize=4.25in
\hspace*{0in}
\epsffile{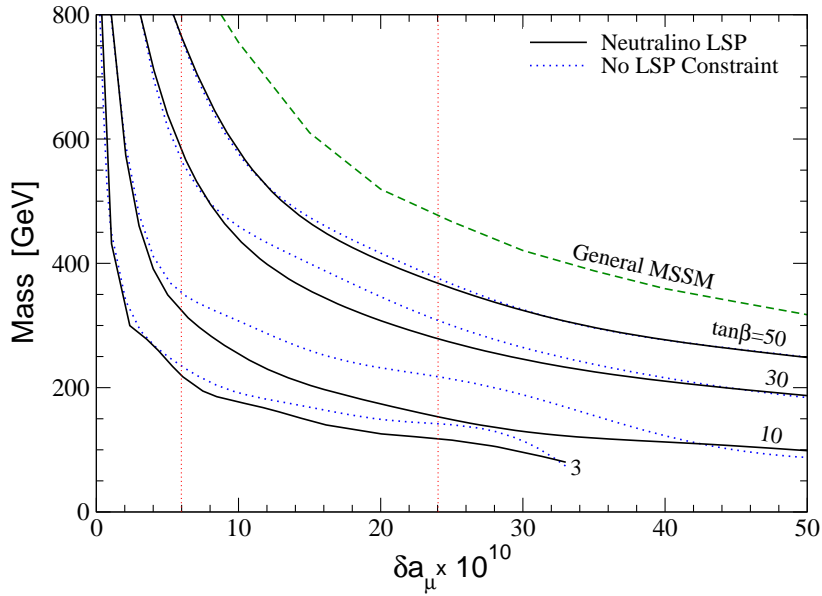}
\caption{Bound on the mass of the LSP as a
function of $\delta a_\mu$ for $\tan\beta=3$, 5, 10, 30 and 50. The 
dotted lines assume gaugino unification only while the solid lines
require additionally that the LSP is a neutralino. The dashed line is
the bound in the general MSSM, calculated at $\tan\beta=50$.}
\label{LSPfig}
\end{figure}

The same plots could be produced with the additional assumption that
the LSP be a neutralino, but we will only show the case for the LSP
bound, in Fig.~\ref{LSPfig}. In this figure, the solid lines
correspond to a neutralino LSP, while the dotted lines are for the
more general case discussed above (\ie, they match the lines in
the $\tan\beta=50$ plot of 
Fig.~\ref{massfig}). Notice that for $\delta a_\mu\gsim 40\times
10^{-10}$ there is little difference between the cases with and
without a neutralino LSP. At the extreme upper and lower
values of $\tan\beta$ there is also little difference. It is only for the
intermediate values of $\tan\beta$ that the mass bound shifts
appreciably; for $\tan\beta=10$ it comes down by as much as $50\gev$
when one imposes a neutralino LSP.

%%%%%%%%%%
%%%%%%%%%%%

\begin{figure}
\centering
\epsfxsize=4.25in
\hspace*{0in}
\epsffile{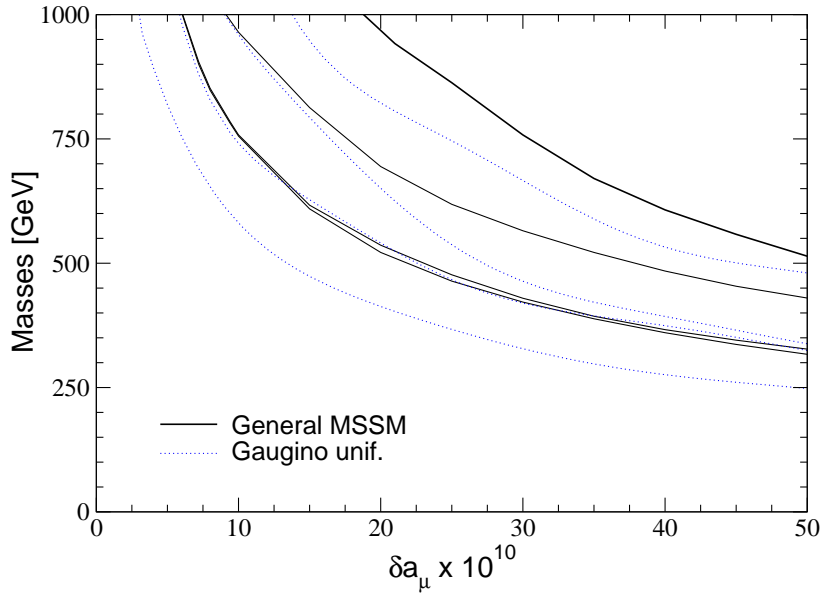}
\caption{Bounds on the masses of the four lightest sparticles as a 
function of $\delta a_\mu$ for $\tan\beta=50$. The 
dotted lines assume gaugino unification only while the solid lines
are for the general MSSM.}
\label{nongutmaxes}
\end{figure}

Finally, we consider the most general MSSM case, \ie,  
without gaugino unification. Here the correlations are much less
pronounced, but interesting bounds still exist. For example the
central value of the E821 data still demands at least 3 sparticles
below $525\,(770)\gev$.
In figure~\ref{nongutmaxes} we demonstrate this explicitly by
plotting the masses of the four lightest sparticles for $\tan\beta=50$
and a wide range of $\delta a_\mu$. (We also plot the mass bound on the LSP
in Fig.~\ref{LSPfig} with the label ``General MSSM.'')
We see that dropping the gaugino
unification requirement has one primary effect: the mass of the LSP is
significantly increased. This is because the LSP in the unified case
is usually a $\neut_1\sim\bino$ but isn't itself responsible for
generating $\delta a_\mu$. In the general case, the LSP must
participate in $\delta a_\mu$ (otherwise its mass could be arbitrarily
large) and so is roughly the mass of the {\it second}\/ lightest
sparticle in the unified case, whether that be a $\smuon$ or $\char$.
Otherwise the differences between the more general MSSM and the
gaugino unified MSSM are small. 

%%%%%%%%%%
%%%%%%%%%%%

We have summarized all this data on the LSP in Table~\ref{masstable}
where we have shown the mass bounds using the $1\sigma$ limit
of the E821 data on the LSP for various $\tan\beta$ values with our various
assumptions. The numbers represent the bounds without (with) the inclusion
of the $\tau$ decay data.  The last line in the table represents an upper
bound for any model with $\tan\beta\leq50$: $m_{\rm LSP}<475\gev\,(1\tev)$
for the E821 $1\sigma$ lower bounds of $\delta a_\mu$.
But perhaps of equal importance are the bounds on the next 2 lightest
sparticles (the ``2LSP'' and ``3LSP''): Using $a_\mu$ calculated
without $\tau$-decay data, $m_{\rm 2LSP}<
485\gev$ and $m_{\rm 3LSP}< 630\gev$; for the value of $a_\mu$
calculated {\it with}\/ the $\tau$ data, $m_{\rm 2LSP}<1\tev$, while 
$m_{\rm 3LSP}$ is pushed somewhat above $1\tev$. Note that the
bound on the 2LSP is essentially identical to that on the LSP.
The central value of the E821 anomaly further implies a fourth (third)
sparticle below $1\tev$ even in this most general case.

%%%%%%%%%%%%%%%%%%%%%%
\begin{table}
\centering
\begin{tabular}{r||c|c|c|}
\multicolumn{1}{c||}{Mass} & General & Gaugino & + Dark \\
\multicolumn{1}{c||}{Bound} & MSSM & Unification & Matter \\ \hline\hline
$\tan\beta=3$ & 205 {\sl (331)}& 140 {\sl (230)}& 115 {\sl (215)}\\
5 & 235 {\sl (395)}& 170 {\sl(280)}& 135 {\sl(280)}\\
10 & 280 {\sl(475)}& 215 {\sl(350)}& 150 {\sl(325)} \\
30 & 340 {\sl (750)}& 305 {\sl(580)}& 275 {\sl(565)}\\
50 &\fbox{{\bf 475} {\sl(1000)}}& {\bf 370} {\sl(765)}& {\bf 365} {\sl(765)}\\
\cline{2-4}
\end{tabular}
\caption{Upper bounds on the mass (in GeV) of the lightest sparticle for the
general MSSM, the MSSM with gaugino mass unification, and the MSSM
with gaugino mass unification plus a neutralino LSP. The entries represent
the $1\sigma$ bound without (with) the inclusion of the $\tau$ decay data.
The boldfaced $\tan\beta=50$ entries represent upper bounds over all
$\tan\beta\leq 50$.}
\label{masstable}
\end{table} 
%%%%%%%%%%%%%%%%%%%%%%%%

\subsection{Bounds on the sparticle species}

In the previous subsection, we derived bounds on the lightest
sparticles, independent of the identity of those sparticles. Another
important piece of information that can be provided by this analysis
is bounds on individual species of sparticles, for example, on the
charginos or on the smuons. These bounds will of 
necessity be higher than those derived in the previous section, but
still provide important information about how and where to look for
SUSY. In particular, they can help us gauge the likelihood
of finding SUSY at Run~II of the Tevatron or at the LHC.

There is one complication in obtaining these bounds. At low
$\tan\beta$ the data is most easily explained by the neutralino
diagrams and as such there must be at least one light smuon and one
light neutralino. At larger $\tan\beta$ contributions from
the chargino diagrams dominate, implying a light chargino and 
sneutrino. However the correlations already discussed preserve the
bounds on the various species over the whole range of $\tan\beta$. A
bound on $m_\sneut$ implies a bound on $m_{\smuon_1}$, and a bound on
$m_{\char_1}$ implies a bound on at least one of the $m_{\neut_i}$,
and in certain cases (such as gaugino unification), 
the converses may be true as well.

\begin{figure}
\centering
\parbox{3truein}{
\epsfxsize=3in
\hspace*{0in}
\epsffile{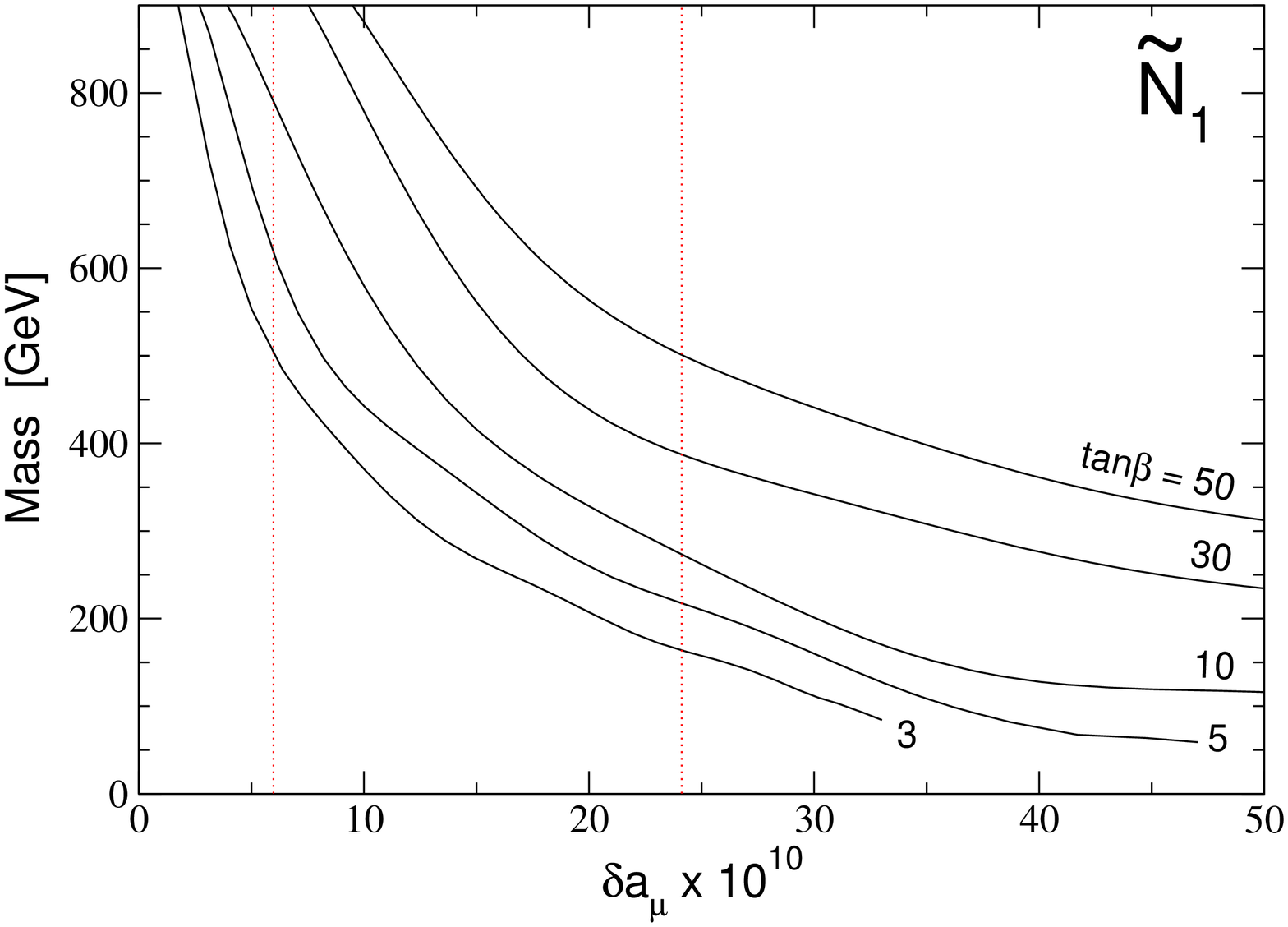}
}~~\parbox{3truein}{
\epsfxsize=3in
\hspace*{0in}
\epsffile{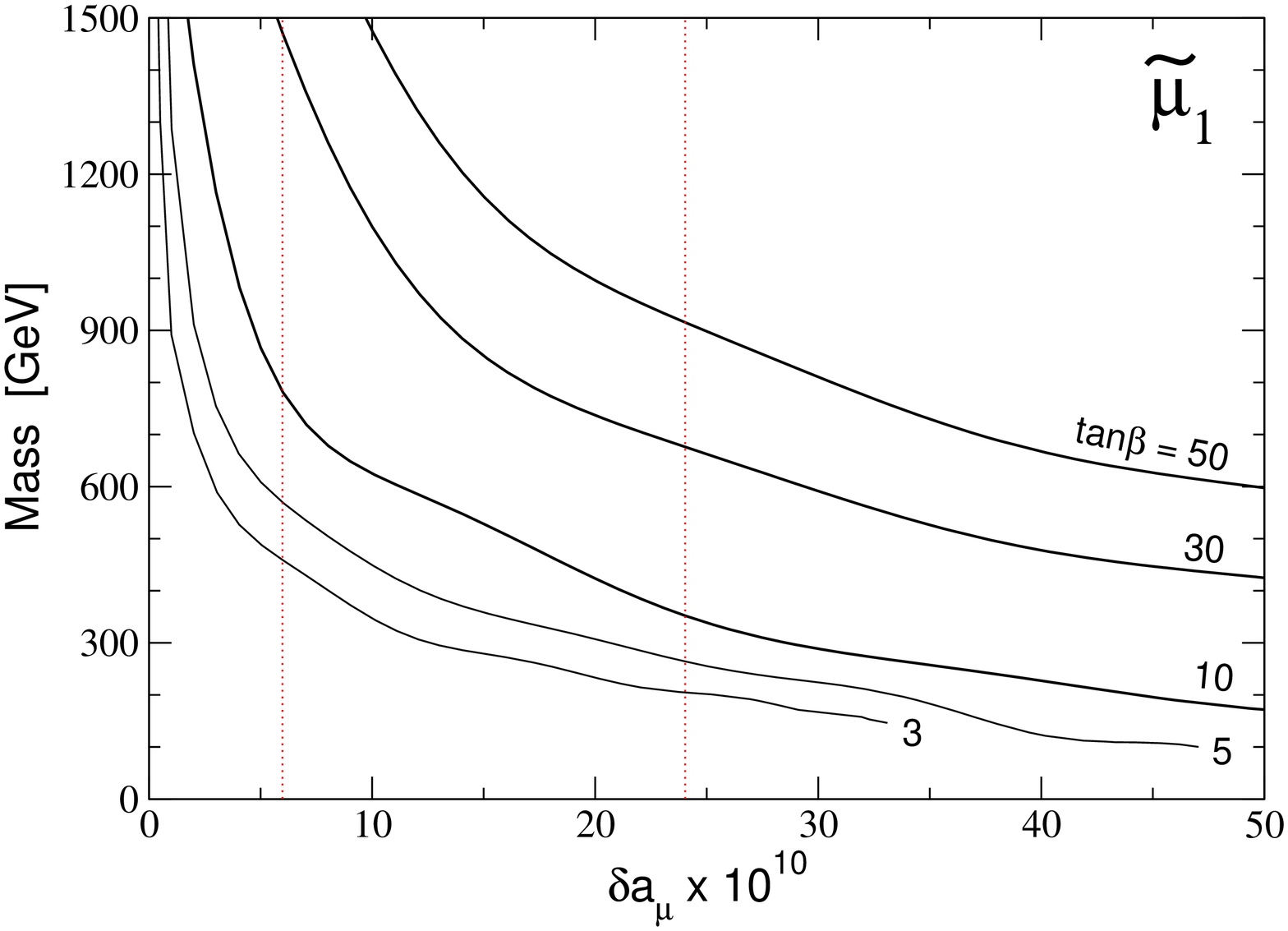}}
\caption{Bounds on the masses of $\neut_1$ and $\smuon_1$
as a function of $\delta a_\mu$ for various $\tan\beta$ with gaugino mass
unification assumed.  Again, the vertical dotted lines indicate $1\sigma$
bounds without (with) the inclusion of the $\tau$ decay data.}
\label{species}
\end{figure}

We have shown in Fig.~\ref{species} the mass bounds on $\smuon_1$ and
$\neut_1$ under the assumption of gaugino unification; a plot for 
$\char_1$/$\neut_2$ will appear later in our discussion of Tevatron
physics. Note that $\neut_1$ must lie below $500\,(950)\gev$, even for large
$\tan\beta$, thanks to the gaugino unification condition, while
$\smuon_1$ can be heavier but must still lie below $915\,(1800)\gev$ at
$1\sigma$.

%%%%%%%%%%
%%%%%%%%%%%

\begin{figure}
\centering
\epsfxsize=4.5in
\hspace*{0in}
\epsffile{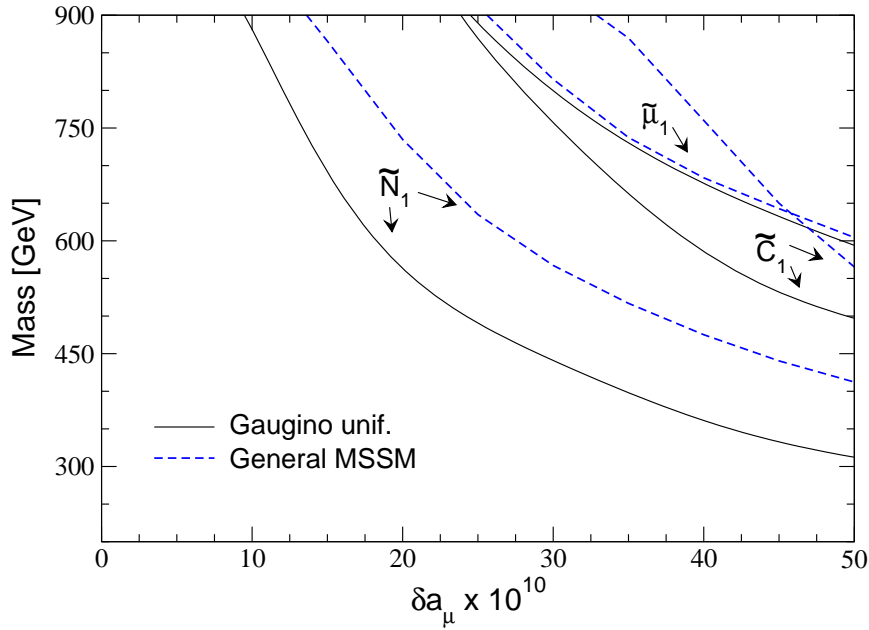}
\caption{Bounds on the masses of $\smuon_1$, $\char_1$,
and $\neut_1$ as a function of $\delta a_\mu$ for $\tan\beta=50$, with
gaugino unification (solid) and in the general MSSM (dashed). The two lines
for the $\smuon_1$ essentially overlap.}
\label{nogut}
\end{figure}

Finally, we can consider the general MSSM without gaugino
unification. The results are shown schematically in Fig.~\ref{nogut}
where the general MSSM bounds (solid lines) are shown alongside the
gaugino unification bounds (dashed lines).

%%%%%%%%%%
%%%%%%%%%%%

We can see from the figure that the bound on the $\smuon_1$ is
essentially identical to that in the gaugino unification
picture. However the gaugino masses have shifted, and the reason is
no mystery. Once again, the
lightest neutralino is no longer a $\bino$-like spectator to the
magnetic moment, but is a $\wino$-like partner of a participating
$\wino$-like chargino.

The results of these plots for the current discrepancy are summarized
as follows. For the MSSM with gaugino unification, the lightest neutralino
must fall below $500\,(950)\gev$ for a $1\sigma$ deviation.
 The lighter smuon must lie below
$915\gev$ ($1.8\tev$). Without the $\tau$-decay data, the lighter
chargino falls below $890\gev$; using the $\tau$ data, the bound rises
above $2\tev$.

%Imposing a neutralino LSP strengthens the bounds on the gaugino sector
%only: now the lightest
%neutralino must lie below $560\gev$ ($400\gev$) and the chargino below
%$1\tev$ ($650\gev$). 

For the worst case, the general MSSM, the smuon bound
increases by only about $50\gev$ over the gaugino unification limits
and the neutralino bounds increase to $890\gev$ (over $1\tev$), only
slightly higher than the unified case. However, the bounds on the
lighter chargino jump above $2\tev$.

%We summarize our results for the various sparticle species in
%Table~\ref{masseachtable}. There we have shown the upper bounds on
%several sparticles in the general MSSM, the MSSM with gaugino
%unification, and the previous case with the additional requirement of
%a neutralino LSP (``dark matter''). The bounds represents those
%obtained using the $1\sigma$ limit (central value) of the E821 data. 

%%%%%%%%%%%%%%%%%%%%%%%
%\begin{table}
%\centering
%\begin{tabular}{c||c|c|c|}
%\multicolumn{1}{c||}{Mass} & General & Gaugino & + Dark \\
%\multicolumn{1}{c||}{Bound} & MSSM & Unification & Matter \\ \hline\hline
%$\neut_1$ & 610 {\sl (455)}& 470 {\sl (350)}& 340 {\sl (265)}\\
%$\neut_2$ & none {\sl (none)}& 830 {\sl(565)}& 575 {\sl(440)}\\
%$\char_1$ &  none {\sl(690)}& 830 {\sl(565)}& 575 {\sl(440)} \\
%$\smuon_1$ & 870{\sl (680)}& 825 {\sl(665)}& 825 {\sl(665)}\\
%$\sneut$ & 865 {\sl(675)}& 820 {\sl(660)}& 820 {\sl(660)}\\
%\cline{2-4}
%\end{tabular}
%\caption{Upper bounds on the mass (in GeV) of various
%sparticles for the
%general MSSM, the MSSM assuming gaugino mass unification, and the MSSM
%with gaugino mass unification plus a neutralino LSP. The entries
%represent the bound for the $1\sigma$ limit {\sl (central value)} of the
%E821 data. These bounds are for all $\tan\beta\leq50$.
%}
%\label{masseachtable}
%\end{table} 
%%%%%%%%%%%%%%%%%%%%%%%%

\subsection{Bounds on $\tan\beta$}

\begin{figure}
\centering
\epsfxsize=4.25in
\hspace*{0in}
\epsffile{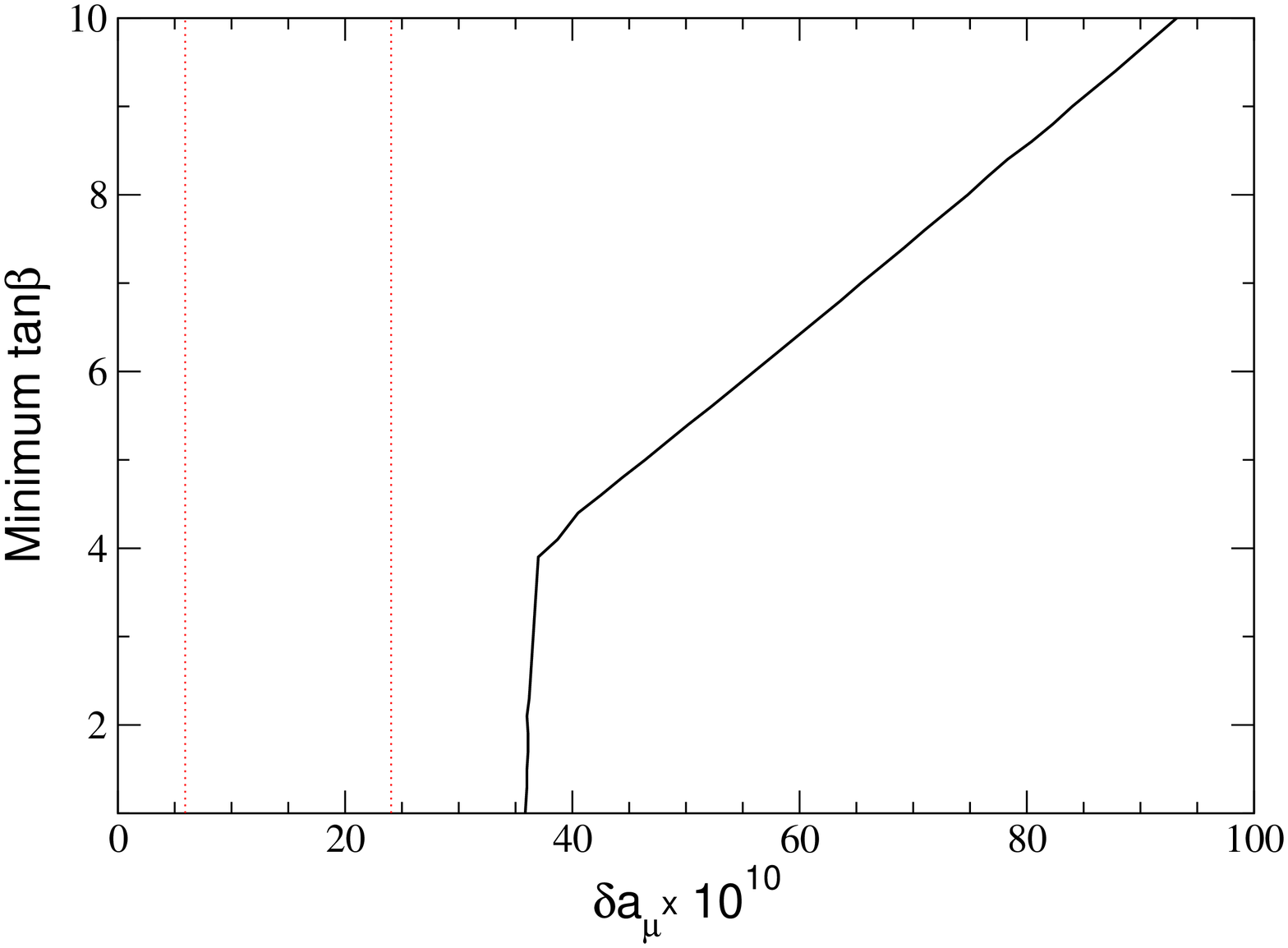}
\caption{Bounds on $\tan\beta$ as a function of $\delta a_\mu$ for the
general MSSM. The dotted lines represent the $1\sigma$ limits, without
(with) the inclusion of the $\tau$ decay data.}
\label{tanb}
\end{figure}

The final bound we will investigate using the E821 data is on
$\tan\beta$. There had been, after the appearance of the original
E821 data, some discussion in the literature about
which values of $\tan\beta$ were capable of explaining it.
In particular, at lower $\tan\beta$ there is a real suppression
in the maximum size of $\delta a_\mu$. In Figure~\ref{tanb} we have
shown the maximum attainable value of $\delta a_\mu$ as a function of
$\tan\beta$ in the general MSSM; adding the assumption of gaugino
unification changes the figure only slightly.

%%%%%%%%%%
%%%%%%%%%%%

The limit in Fig.~\ref{tanb} clearly divides into two regions. At
$\delta a_\mu>36\times 10^{-10}$ the chargino contribution
dominates and thus $\delta a_\mu\propto y_\mu$, scaling
linearly with $\tan\beta$. At lower $\delta a_\mu$, however, both
neutralino and chargino contributions can be important so it becomes
possible to generate $\delta a_\mu$ with much smaller values of
$\tan\beta$ than would be possible from the charginos alone. The E821
data does not therefore imply any bound on $\tan\beta$ whatsoever,
neither at $1\sigma$ nor at the experimental central value. Further
reductions in the size of the error bars will not change this result,
so long as the central value remains at or below its current $1\sigma$
upper bound.

\subsection{Implications for the Tevatron}

At its simplest level, the measurement of $\delta a_\mu$, an anomaly
in the lepton sector, has little impact on the Tevatron, a hadron
machine. In particular, the light smuons associated with $\delta
a_\mu$ cannot be directly produced at the Tevatron, occurring only if
heavier non-leptonic states are produced which then decay to
sleptons. In the calculation of $a_\mu$, the only such sparticles are
the neutralinos and charginos. These states can be copiously produced
and in fact form the initial state for the ``gold-plated'' SUSY
trilepton signature. 

Of particular interest for the trilepton signature are the masses of
the lighter chargino ($\char_1$) and 2nd lightest neutralino ($\neut_2$).
Studies of mSUGRA parameter space indicate that the sensitivity to the
trilepton signature at Run II/III of the Tevatron depends strongly on
the mass of sleptons which can appear in the gaugino decay chains. For
heavy sleptons, the Tevatron is only sensitive to gaugino masses in
the range~\cite{tevsugra}
$m_{\char_1,\neut_2}\lsim 130$ to $140\gev$ for $10\invfb$ of 
luminosity and $145$ to $155\gev$ for $30\invfb$, where the quoted
ranges take one from low to high $\tan\beta$.
However, for light sleptons (below about
$200\gev$) the range is considerably extended, up to gaugino masses
around 190 to $210\gev$. 
%Of course, this latter range is exactly the
%one most relevant for understanding $a_\mu$.

\begin{figure}
\centering
\epsfxsize=4.25in
\hspace*{0in}
\epsffile{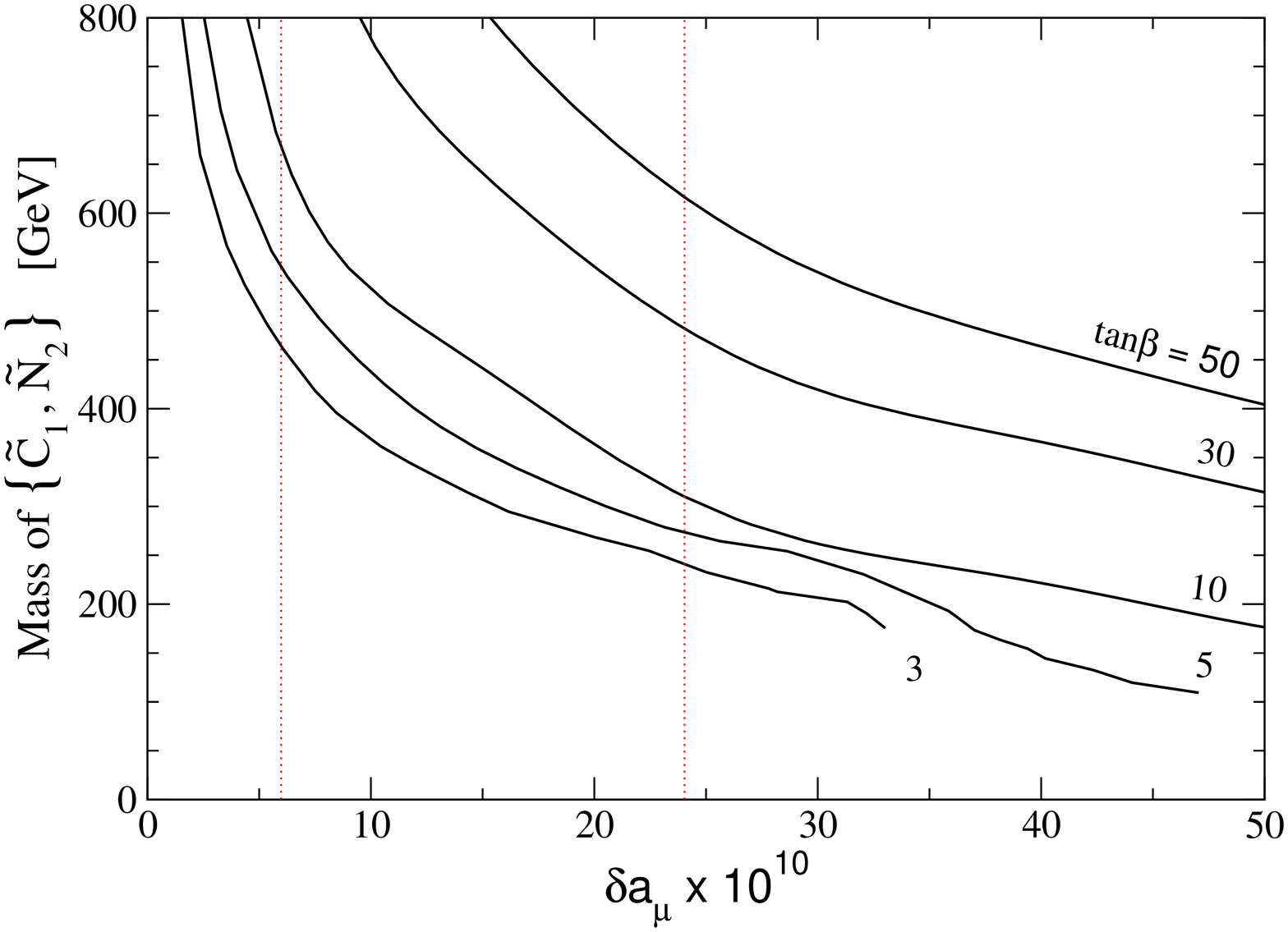}
\caption{Mass bounds on $\char_1$ and $\neut_2$ (where
$m_{\char_1}\simeq m_{\neut_2}$) as a function of
$\delta a_\mu$ for $\tan\beta=3$, 5, 10, 30 and 50. The dotted lines
represent the $1\sigma$ bounds without (with) the inclusion of the $\tau$
decay data. This figure assumes gaugino unification and a neutralino LSP.}
\label{tevatron}
\end{figure}

It is impossible in the kind of analysis presented here
to comment on the expected
cross-sections for the neutralino-chargino production (for example, 
there is
no information in $a_\mu$ on the masses of the $t$-channel squarks)
but we can examine the mass bounds on $\char_1$ and $\neut_2$. In
Fig.~\ref{tevatron} we have shown just that: the upper bound on the
{\em heavier}\/ of either $\char_1$ or $\neut_2$ as a function of
$\delta a_\mu$ for several values of $\tan\beta$. 

A few comments are in order on the figure. First, this figure assumes
gaugino unification; dropping that assumption can lead to
significantly heavier and unequal masses for the $\neut_2$ and $\char_1$.
Second, we have also assumed a neutralino LSP; this is to be
expected since the event topology for the trilepton signal assumes a
stable, neutralino LSP. Finally, on the $y$-axis is actually plotted
$m_{\char_1}$, but in every case we examined with gaugino unification,
the difference in the maximum masses of $\char_1$ and $\neut_2$
differed by at most a few GeV. This is because they are both
dominantly wino-like in the unified case and thus have masses
$\simeq M_2$.

%%%%%%%%%%
%%%%%%%%%%%

From the figure it is clear that one cannot devise a no-lose theorem for the
Tevatron from current E821 data. However, if $\tan\beta$ is small and 
$\delta a_{\mu} \ge 35 \times 10^{-10}$
(the current central value using only $e^+e^-$ data)
then the gauginos may be within the Tevatron's reach.  We must also 
emphasize that these are {\em upper bounds}\/ on the
sparticle masses and in no way represent best fits or preferred
values. Thus, for small $\tan\beta$,
there is reason to hope that the Tevatron will be able to probe
the gaugino sector in Run II or III; for larger $\tan\beta$ there is
little information about SUSY at the Tevatron to be garnered from
$a_\mu$.

\subsection{Implications for a Linear Collider}

A consensus has emerged in favor of building a
$\sqrt{s}=500\gev$ linear collider, presumably a factory for
sparticles with masses below $250\gev$. What does the measurement of
$a_\mu$ tell us with regards to our chances for seeing SUSY at
$\sqrt{s}=500\gev$? And how many sparticles will be actually
accessible to such a collider? 

\begin{figure}
\centering
\epsfxsize=4.25in
\hspace*{0in}
\epsffile{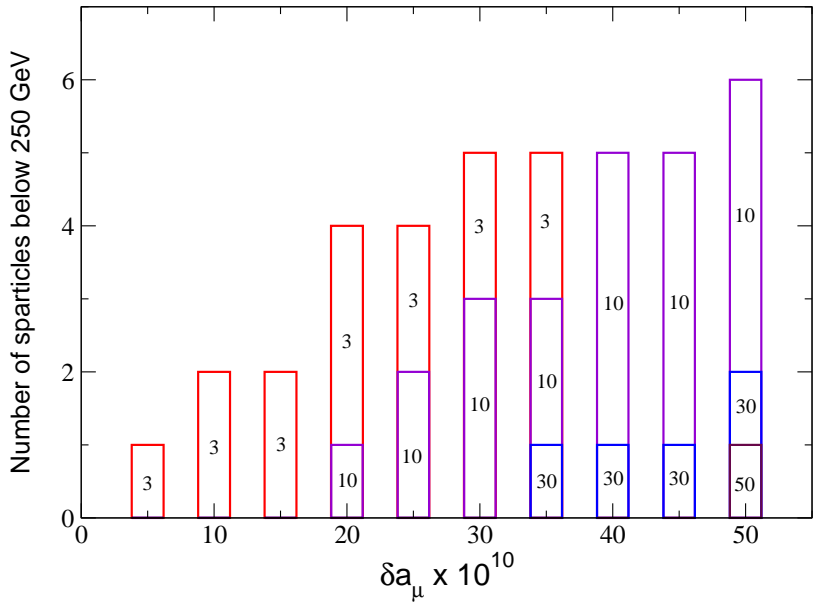}
\caption{Minimum number of sparticles directly observable at a
$\sqrt{s}=500\gev$ linear collider as a function of $\delta a_\mu$. 
The bars represent $\tan\beta=3$, 10, 30 and 50. This graph assumes
gaugino unification but does not include additional sleptons
implied by slepton mass universality.
%The bins in $\delta a_\mu$ are centered on 
%$\delta a_\mu=(15,25,35,\cdots)\times 10^{-10}$.
}
\label{nlchist}
\end{figure}

The analysis of the previous section
can put a lower bound on the number of observable sparticles at a
linear collider as a function of $\delta a_\mu$ and $\tan\beta$ and we
show those numbers as a histogram in Fig.~\ref{nlchist}. In this
figure, we have shown the {\em minimum}\/ number of sparticles with
mass below $250\gev$ for $\tan\beta=3$, 10, 30 and 50, assuming
gaugino unification. In the graph,
the thinner bars represent smaller $\tan\beta$. As is to be
expected, the number of light states increases with increasing
$\delta a_\mu$ and with decreasing $\tan\beta$. However note
that there are no $\tan\beta=3$ lines for $\delta a_\mu\geq 40\times
10^{-10}$ since there is no way to explain such large $\delta a_\mu$
values at low $\tan\beta$.

%%%%%%%%%%
%%%%%%%%%%%

We see from the figure that at low $\tan\beta$, there is a $1\sigma$
``guarantee'' that at least 1 to 4 sparticles will be produced at a
$500\gev$ linear collider, depending on whether the $\tau$-decay is
used. This counting does not include extra sleptons due to slepton
mass universality; for example, a light muon sneutrino also implies
light tau and electron sneutrinos, and likewise for the charged smuon. 
We see also that for $\tan\beta\gsim30$ there is no such guarantee that a
$500\gev$ machine would produce on-shell sparticles; this is not to be
taken to mean that one should not expect their production, simply that
$a_\mu$ cannot guarantee it. 

A similar bar graph can also be made for a $1\tev$ machine, though we
do not show it here. However the relevant numbers can be inferred from
Fig.~\ref{massfig}.
%we see that such a machine is ``guaranteed'' (at
%$1\sigma$) to produce at least four sparticles for low $\tan\beta$ and
%at least 1 sparticle for $\tan\beta\lsim 30$; there is no such 
%guarantee of sparticle production at $\tan\beta> 30$.

Note that any particular class of models, such as gravity- or gauge-mediated
SUSY-breaking, may not saturate the bounds presented here. That is, other
constraints may rule out all regions of
parameter space in which $\delta a_\mu$ exceeds some maximum
value. However if a model does allow a particular value of $\delta
a_\mu$, then pursuant to the conditions discussed here, that model
must have at least as many light particles as the number given above.

\section{Conclusions}

Deviations in the muon anomalous magnetic moment have long been
advertised as a key hunting ground for indirect signatures of SUSY. 
However, the current experimental excess is too small, given the large
theoretical uncertainties, to provide statistically significant
evidence for SUSY. 

The most recent Standard Model calculations indicate an excess 
of either $1.5\sigma$ or $3.2\sigma$. 
If one were to accept the E821 anomaly as evidence for SUSY 
(at either of these values),
then a number of statements can be inferred at the $1\sigma$
confidence level:
\begin{itemize}
\item The lightest sparticle must lie below $475\gev \,(1\tev)$;
\item For models with unified gaugino masses, 
there must be at least 2 sparticles with masses below $585\gev \,(1\tev)$;
\item There is no lower bound on $\tan\beta$.
\end{itemize}
Bounds on individual species of sparticles are weaker, usually
falling at or above $1\tev$. 

However, these pessimistic bounds are over all $\tan\beta$, which
means effectively that they are the bounds when $\tan\beta=50$, the
maximum value we considered. For low $\tan\beta$, the bounds on
sparticle masses are much smaller. At $\tan\beta=3$ and with gaugino
unification and a neutralino LSP, there must be a sparticle below
$115\,(215)\gev$, within the range accessible in the very near future. So,
while the bounds placed on SUSY by $a_\mu$ are relatively weak when no
constraints are placed on the MSSM, constraining the model by
demanding gaugino unification or low(er) $\tan\beta$ can bring down
the mass bounds into the experimentally interesting region.

\section*{Acknowledgments}

We are grateful to S.~Martin for discussions during the early parts of
this project. CK would also like to thank the Aspen Center for Physics
where parts of this work were completed, and the Notre Dame
High-Performance Computing Cluster for much-needed computing
resources. This work was supported in part by the National Science
Foundation under grant NSF-0098791. The work of MB was supported in
part by a Notre Dame Center for Applied Mathematics Graduate Summer
Fellowship.

\section*{Appendix}

The supersymmetric contributions to $a_\mu$ are generated by diagrams
involving charged smuons with neutralinos, and sneutrinos with
charginos. The most general form of the calculation, including phases,
takes the form (we follow Ref.~\cite{martinwells}):
\bea
\delta a_\mu^{(\neut)} & = & \frac{m_\mu}{16\pi^2}
   \sum_{i,m}\left\{ -\frac{m_\mu}{ 12 m^2_{\smuon_m}}
  (|n^L_{im}|^2+ |n^R_{im}|^2)F^N_1(x_{im}) 
 +\frac{m_{\neut_i}}{3 m^2_{\smuon_m}}
    {\rm Re}[n^L_{im}n^R_{im}] F^N_2(x_{im})\right\}%\phantom{xxxx}
\nonumber \\
\delta a_{\mu}^{(\char)} & = & \frac{m_\mu}{16\pi^2}\sum_k
  \left\{ \frac{m_\mu}{ 12 m^2_{\sneut}}
   (|c^L_k|^2+ |c^R_k|^2)F^C_1(x_k)
 +\frac{2m_{\char_k}}{3m^2_{\sneut}}
         {\rm Re}[ c^L_kc^R_k] F^C_2(x_k)\right\}%\phantom{xxxx}
\eea
where $i=1,2,3,4$, $m=1,2$, and $k=1,2$ label the neutralino, smuon
and chargino mass eigenstates respectively, and
\bea
n^R_{im} & = &  \sqrt{2} g_1 N_{i1} X_{m2} + y_\mu N_{i3} X_{m1},
\nonumber \\
n^L_{im} & = &  {1\over \sqrt{2}} \left (g_2 N_{i2} + g_1 N_{i1}
\right ) X_{m1}^* - y_\mu N_{i3} X^*_{m2}, \nonumber \\
c^R_k & = & y_\mu U_{k2}, \nonumber \\
c^L_k & = & -g_2V_{k1},
\eea
$y_\mu = g_2 m_\mu/\sqrt{2} m_W \cos\beta$ is the muon Yukawa
coupling, and $g_{1,2}$ are the U(1) hypercharge and SU(2) gauge couplings.
The loop functions depend on the
variables $x_{im}=m^2_{\neut_i}/m^2_{\smuon_m}$, 
$x_k=m^2_{\char_k}/m^2_{\sneut}$ and are given by
\bea
F^N_1(x) & = &\frac{2}{(1-x)^4}\left[ 1-6x+3x^2+2x^3-6x^2\ln x\right] 
\nonumber \\
F^N_2(x) & = &\frac{3}{(1-x)^3}\left[ 1-x^2+2x\ln x\right] 
\nonumber \\
F^C_1(x) & = &\frac{2}{(1-x)^4}\left[ 2+ 3x - 6
x^2 + x^3 +6x\ln x\right] \nonumber \\
F^C_2(x) & = &-\frac{3}{2(1-x)^3}\left[ 3-4x+x^2 +2\ln x\right]
\eea
For degenerate sparticles ($x=1$) the functions are normalized so that
$F^N_1(1) = F^N_2(1) = F^C_1(1) = F^C_2(1) = 1$. We can also bound the
magnitude of some of these functions; in particular $|F^N_2(x)|\leq 3$
while $|F^C_2(x)|$ is unbounded as $x\to0$.

The neutralino and chargino mass matrices are given by
\beq
M_{\neut} = \left(\begin{array}{cccc} M_1 & 0 & - m_Z s_W c_\beta &
m_Z s_W s_\beta \\
0 & M_2 & m_Z c_W c_\beta & - m_Z c_W s_\beta \\
-m_Z s_W c_\beta & m_Z c_W c_\beta & 0 & -\mu \\
m_Z s_W s_\beta & - m_Z c_W s_\beta& -\mu & 0 \end{array} \right)
\label{neutralinomassmatrix} 
\eeq
and
\beq
M_{\char} = 
\left(\begin{array}{cc} M_2 & \sqrt{2} m_W s_\beta \\
                              \sqrt{2} m_W c_\beta & \mu \end{array}
	\right)
\label{charginomassmatrix}
\eeq
where $s_\beta = \sin\beta$,
$c_\beta = \cos\beta$ and likewise for $\theta_W$.
The
neutralino
mixing matrix 
$N_{ij}$ 
and the chargino mixing  matrices $U_{kl}$ and $V_{kl}$ satisfy
\bea
N^* M_{\neut} N^\dagger &=& {\rm diag}(
m_{\neut_1},
m_{\neut_2},
m_{\neut_3},
m_{\neut_4}) \nonumber\\
U^* {M}_{\char} V^\dagger &=& {\rm diag}(
m_{\char_1},
m_{\char_2})
. 
\eea

The smuon mass matrix is given in the 
$\{ \smuon_L, \smuon_R \}$ basis as:
\beq
M^2_{\smuon}=\left( \begin{array}{cc}
	M_L^2 & m_\mu (A^*_{\tilde\mu}-\mu\tan\beta) \\
	m_\mu (A_{\tilde\mu}-\mu^*\tan\beta) & M_R^2 \end{array}
	\right)
\eeq
where
\bea
M_L^2 & = & m^2_L +(s_W^2 -\frac{1}{2})m_Z^2\cos 2\beta \nonumber
\\
M_R^2 & = &  m^2_R -s_W^2 \, m_Z^2\cos 2\beta 
\eea
for soft masses $m_L^2$ and $m_R^2$;
the unitary smuon mixing matrix $X_{mn}$ is defined by
\bea
X M^2_{\smuon}\, X^\dagger = 
{\rm diag}\, (m^2_{\smuon_1}, m^2_{\smuon_2}).
\eea
We will define a smuon mixing angle $\theta_\smuon$ such that 
$X_{11}=\cos\theta_\smuon$ and $X_{12}=\sin\theta_\smuon$. In our
numerical calculations we will set $A_\smuon=0$. At low $\tan\beta$ we
have checked that varying $A_\smuon$ 
makes only slight numerical difference, while at large
$\tan\beta$ it has no observable effect whatsoever. Finally, 
the muon sneutrino mass is related to the left-handed smuon mass parameter
by
\beq
m_{\sneut}^2 = m_L^2 + {1\over 2} m_Z^2 \cos 2\beta  .
\eeq

The leading 2-loop contributions to $\delta a_\mu$ have been
calculated~\cite{2loop} and have been found to suppress the SUSY
contribution by a factor $(4\alpha/\pi)\log(m_{\rm
SUSY}/m_\mu)\approx 0.07$; we will include this 7\% suppression in all
our numerical results.

\end{document}